\documentclass[aps,prc,preprint,superscriptaddress]{revtex4}

\usepackage{graphics}

\begin{document}

\title{Verification of the Surrogate Ratio Method}

\author{Satoshi CHIBA}
\email{chiba.satoshi@jaea.go.jp}
\affiliation{Japan Atomic Energy Agency, Tokai, Naka, Ibaraki 319-1195, Japan}
\affiliation{National Astronomical Observatory of Japan, Mitaka, Tokyo 181-8588, Japan}

\author{Osamu IWAMOTO}
\affiliation{Japan Atomic Energy Agency, Tokai, Naka, Ibaraki 319-1195, Japan}

\date{\today}

\begin{abstract}

Effects of difference in the spin and parity distributions for the surrogate
and neutron-induced reactions are investigated.  
Without assuming specific (schematic) spin-parity distributions, 
it was found that the surrogate ratio method can be employed to
determine neutron fission and capture
cross sections if 1) weak Weisskopf-Ewing 
condition (defined in this paper) is satisfied, 
2) there exist two surrogate reactions whose spin-parity 
distributions of the decaying nuclei are almost equivalent,
and 
3) difference of the representative spin
values between the neutron-induced and surrogate reactions is 
no much larger than 10 $\hbar$.  
If these conditions are satisfied, we need not to know the 
spin-parity distributions populated by the surrogate method.
Instead, we should just select a pair of surrogate 
reactions which will populate the 
similar spin-parity distributions, using targets having similar structure and
reactions having the similar reaction mechanisms.  
Achievable accuracy is
estimated to be around 5 and 10 \% for fission and capture channels,
respectively, for nuclei of the Uranium region.  
The surrogate absolute method, on the contrary, can be marginally
applicable to determination of fission cross sections. However, there will be 
little hope
to apply this method for capture cross section measurements unless the
spin-parity distributions in the neutron-induced and surrogate reactions
are fairly close to each other or the difference can be corrected 
theoretically.
The surrogate ratio method was shown also to be a robust method in the
presence of breakup reactions, again, without assuming specific
breakup reaction mechanisms.
\end{abstract}


\maketitle

\section{Introduction}

With the advance of nuclear science and technology, neutron cross sections of
unstable nuclei, such as minor actinides (MAs) and 
long-lived fission products (LLFPs), are becoming more and more
necessitated. 
Neutron cross sections of radioactive nuclei also play important roles in
astrophysical nucleosynthesis.  
In spite of the importance, however, measurement of neutron
cross sections are extremely difficult for these nuclei since preparation of
enough amount of sample is difficult or practically impossible. At the same
time, theoretical determination of the fission and capture cross sections
still suffers from a large uncertainty if there exists no experimental data; an
error of factor of 2, namely the uncertainty of 100 \%, will be a reasonable
estimate. These fundamental problems prevent us from accurate
determination of neutron cross sections of unstable nuclei 
including MAs and LLFPs.

Recently, a new method, called surrogate method, has come to be used actively 
to
determine neutron cross sections of unstable nuclei (see, e.g., 
Refs. \cite{cramer70,britt79,petit04,boyer06,younes03a,younes03b,escher06,lyles07,forssen07,lesher09,allmond09}
and references therein). This is a method which uses (multi) nucleon transfer
reactions (both stripping and pick-up) or 
inelastic scattering on available target nuclei and produce the
same compound nuclei as those of 
the desired neutron-induced reactions, and measure
the decay branching ratios leading to capture and/or fission channel.
Identification of the produced compound nuclei and their excitation energies 
can be done by detection of
the ejectile species and their energies.

At a first glance, it seems to be a simple and effective method to simulate
the neutron-induced reactions.
However, the thing is not that easy. Even if we produce the same
compound nuclei at the same 
excitation energy as
produced in the desired neutron-induced reactions, the 
spin-parity distributions are plausibly
different between them. 
Since we are interested in low-energy neutron cross sections relevant to
reactor applications and astrophysics, the produced 
compound nuclei decay statistically, and the branching ratio is strongly
influenced by the spin and parity. Therefore, difference of the spin-parity
distributions between the surrogate and neutron-induced reactions must be
properly taken into account in
converting the branching ratio determined by the surrogate method to the one
for neutron-induced reactions. Up to now, however, it has not been able to
deduce the spin-parity distribution in the surrogate reactions, since they are
normally multi-nucleon transfer reactions, the reaction mechanisms of
which are not understood well. What have been done so far is to assume that the
decay branching ratio does not depend on the spin-parity and ignore the
difference; the so-called Weisskopf-Ewing condition, or
to assume schematic (rather arbitrary) 
spin-parity distributions for the surrogate reaction and
argue that they do not affect the decay branching ratio sensitively. Both of
these approaches, however, are based on 
arbitrary assumptions which have not been
justified theoretically nor experimentally. 
On the other hand, it is also true that the surrogate method has
yielded a rather accurate cross sections, verified when the
corresponding neutron data are available. Therefore, it is natural to expect
that there is a certain condition to equate the results from the surrogate
method and the neutron-induced reactions. However, the condition under which 
the surrogate method works is not clearly understood yet.

In this paper, we investigate the spin-parity dependence of the branching
ratios of Uranium isotopes to the fission and capture channels and clarify
the condition for the surrogate (ratio) method to work, and
estimate the accuracy achievable by it.

\section{Surrogate ``absolute" and ``ratio" methods}

In the surrogate method, we measure a branching ratio to a specific decay
channel, normally the fission or capture channel by populating the same kind
of compound nucleus as the desired neutron-induced reactions . 
We denote the decay channel
by a subscript $i$ ($i$ = fission or capture), and then the surrogate method
hopefully gives a ratio of the neutron cross section $\sigma^n_i$ 
to the total neutron
reaction cross section $\sigma^n_{R}$ of the compound system, namely,
\begin{equation}
\label{eq:a}
R^S_{i}\stackrel{?}{=}\frac{\sigma^n_{i}}{\sigma^n_{R}},
\end{equation}
%
The symbol $R^S_i$ denotes the branching ratio of the nucleus decaying 
to channel $i$ 
populated by the surrogate reaction, and is defined later by Eq. (\ref{eq:2a}).
By multiplying it the total reaction cross section $\sigma^n_{R}$ calculated
by the optical or coupled-channel model, we can determine the 
neutron cross section $\sigma^n_{i}$.  Here, a question mark is explicitly
shown since it is not obvious if this equality holds or not. 
It is due to the reason that
the spin-parity distributions populated in the surrogate (left-hand-side) and 
neutron-induced (right-hand-side) reactions
are different, and the branching ratio is dependent on them in general.
This is the very fundamental problem to be resolved for the surrogate method
to yield correct neutron-indeed cross sections.
This method is referred to as the surrogate absolute method. On
the contrary, these ratios can be measured for two nearby
nuclei 1 and 2 by using the same kind of surrogate reactions, $S_{1}$ and
$S_{2}$, e.g., ($t,p$) reactions on different targets. If we know the neutron
cross section $\sigma_{i}^{n_{2}}$ for the reaction leading
to the same compound nucleus as the $S_{2}$ reaction, 
we can determine the neutron
cross section ($\sigma_{i}^{n_{1}}$) which leads to the same
compound nucleus as $S_{1}$ reaction via the equality
(with a question mark)
\begin{eqnarray}
\label{eq:b}
\frac{R_{i}^{S_{1}}}{R_{i}^{S_{2}}}  &\stackrel{?}{=}& \frac{\frac{\sigma_{i}^{n_{1}}}
{\sigma_{R}^{n_1}}}{\frac{\sigma_{i}^{n_{2}}}{\sigma_{R}^{n_2}}}, \\
\rightarrow \sigma_{i}^{n_{1}}  &\stackrel{?}{=}&
 \sigma_{i}^{n_{2}}\cdot
\frac{\sigma_{R}^{n_{1}}}{\sigma_{R}^{n_{2}}}
\cdot\frac{R_{i}^{S_1}}{R_{i}^{S_2}
},~~(i=\mathrm{fission~or~capture}).
\end{eqnarray}
Here, $\sigma_{i}^{n_{j}}$ denotes the neutron fission ($i$=fission) or
capture ($i$=capture) reaction cross section, and $\sigma_{R}^{n_{j}}$ the
total neutron reaction cross section for the reaction $n_{j}$ ($j$ = 1 or 2). 
Provided that the above equations hold, we can determine the neutron cross
section $\sigma_{i}^{n_{1}}$ from this formula, 
since we know $\sigma_{i}^{n_{2}}$, 
we measure the ratio $R_{i}^{S_1}/R_{i}^{S_2}$
and we can calculate the ratio of the reaction cross sections $\sigma
_{R}^{n_{1}}/\sigma_{R}^{n_{2}}$ by the coupled-channel theory rather
accurately\cite{capote08,kunieda09}. This method is referred to as
the surrogate ratio method or relative surrogate
method. 
It is naively expected to give a result better than the surrogate
absolute method, 
since we do not need to know in the relative method 
all the experimental artifacts such as the 
detector efficiency and geometrical factor required to deduce
the ratio in the absolute method.
However, all these methods require a fact
that the branching ratios are equal for the surrogate and the neutron-induced
reactions. This is true only when 1)
the ratios are independent of the spin-parity of the decaying nuclei
(Weisskopf-Ewing condition\cite{we}), or 2) the spin-parity distributions are
equivalent for the surrogate and neutron reactions, or 3) the ratio is not
sensitive to the difference of the spin-parity distributions between the 
neutron-induced and surrogate reactions. Below, we will
investigate if these assumptions are justified or not, and when justified,
what accuracy will be.  

\section{Computational method and results}

We use the Hauser-Feshbach theory\cite{hf} to calculate the decay branching
ratios of various spin-parity ($J^{\pi}$) states of $^{239}$U by using
CCONE code system\cite{ccone}. It represents a nucleus
produced by n+$^{238}$U reactions and corresponding surrogate reactions such
as $^{237}$U($t,p$)$^{239}$U. This nucleus was chosen just as an example. In
the calculation, the same parameter values for 
discrete level structures, transmission coefficients, level
density, fission barrier and GDR as used in the
evaluation of JENDL Actinoid File 2008\cite{jac08} were used. Therefore, the
present calculation contains realistic information of the characteristics of
participating nuclei adjusted to reproduce neutron cross sections.

Figures 1 and 2 shows the branching ratios (decay probabilities) 
to the fission (Fig. 1) and capture
(Fig. 2) channels for various $J^{\pi}$ states of $^{239}$U up to $J^{\pi}$ =
(21/2)$^{\pm}$ and neutron energy of 5 MeV. The upper panels in Figs. 1 and 2
show branching ratios from positive parity states, while
the lower ones denote those from negative parity states. If the
Weisskopf-Ewing condition is fulfilled, the various lines in these figures
must coincide (at least approximately); if it is the case both of the
surrogate absolute and ratio 
methods can be justified. However, Fig. 1 shows that the fission
decay ratio varies depending on $J^{\pi}$ by about 15 \% at 5 MeV but
variation is about 50 \% at 1.5 MeV. The convergence is much worse for the
capture channel as shown in Fig. 2; the branching ratios scatter by a factor
of about 10 at 5 MeV, and the variation is much larger at lower energies.
Therefore, we have to conclude that there is only little hope 
to use surrogate method to
determine neutron capture cross sections at these energies, since the
low-energy 
neutron-induced reactions bring only small angular momentum to the compound
system in general, while the surrogate method will bring much more. 
The absolute
surrogate method, therefore, will never work to measure capture cross
sections unless the
spin-parity distribution between the neutron-induced and surrogate reactions
are fairly close to each other or the difference is corrected 
theoretically.  It will be also only marginally applicable to measure the
fission cross sections.

However, the $J^{\pi}$ dependence of the branching ratios to the fission and
capture reactions show rather systematic behaviors. Above 2.5 MeV, the
fission probability shown in Fig. 1 increases monotonically as \textit{J}
increases. Same trend is true for the capture reaction. Since it was found 
also to be true
for other compound nuclei in this mass region, 
$^{236}$U and $^{237}$U (not shown here), 
we may expect that there
is a possibility to cancel out the large $J^{\pi}$ dependence by taking ratios
of the branching ratios for each $J^{\pi}$.
We have done such calculations and the results are shown in Figs. 3 and 4.
Figure 3 shows the ratios of fission probabilities (branching ratios) for
$^{239}$U and $^{237}$U for various values of $J^{\pi}$. We can notice an
astonishingly good convergence. The
thick black line denotes the ratios of the neutron fission probabilities
($\sigma^n_f/\sigma^n_R$) for the corresponding neutron-induced
reactions.   All the curves converges to the ratio of the neutron 
fission probabilities very well. The deviation
is only a level of 3 \% at 5 MeV. The largest scatter lies at about 1.6 MeV,
but the scatter around the neutron curve is only a level of several \% 
nominally, 
while that
was about 50 \% in Fig. 1. This means that we can determine the unknown
fission cross sections by taking this kind of ratio if we know
one of the other neutron cross section. 
The convergence seems to be valid also for somewhat higher value of spins.
Similar convergence, although less dramatic, can be seen in Fig. 4
for capture probabilities. At 5 MeV, the ratios of the capture branching
ratios for the 2 nuclei scatter only by about $\pm$ 5 \% around those for the
neutron capture reaction. At energies from 2.5 to 4 MeV, the surrogate ratios
are all larger than the neutron ratio, but the deviation is still only 10 \%.
The same ratios were compared for various $J^+$ states produced in the 
neutron-induced reactions on $^{197}$Au and $^{193}$Ir in Fig. 5.  
We can notice that very good mutual convergence 
up to 8$^+$ and equivalence to the 
neutron ratio are obtained in this mass region as well.
Therefore, these data can be used to determine the 
GDR parameters at an energy region of, e.g., 
2 to 5 MeV to normalize the calculated neutron capture cross section, 
and these parameters can be used to calculate the capture cross sections at
lower energies since the
Hauser-Feshbach theory can predict the shape of the energy dependent
cross section rather accurately 
if normalization is given correctly at certain energies.
Therefore, there is a fair possibility that we can determine the neutron
capture cross section with accuracy of several \% by the surrogate
ratio method in combination with a theoretical calculation. 
The convergence of the ratios of fission and
capture probabilities are very important to validate the surrogate technique
and can be a base of the validity of surrogate ratio method.

\section{Formal verification of the surrogate ratio method}

In the previous section, we have seen that the ratios of fission and capture
probabilities at various values of $J^\pi$ 
between 2 nuclei have a dramatic convergence
to the ratios of the neutron reactions. This can be 
utilized to verify the surrogate ratio method as follows. Let
2 surrogate reactions used for the ratio method be denoted
as \textit{S}$_{1}$ and \textit{S}$_{2}$, and corresponding neutron reactions
as \textit{n}$_{1}$ and \textit{n}$_{2}$. The reactions \textit{S}$_{j}$ and
\textit{n}$_{j}$ (\textit{j}=1,2) are chosen to lead to the same compound
nucleus. Let us assume that we know the neutron cross section $\sigma
_{i}^{n_{2}}$ for the $n_{2}$ reaction . The branching ratio of the surrogate
reaction for channel $i$ ($i$ = fission or capture) may be written as
$B_{i}^{S_{j}}(U,J^{\pi})$, where $U$ denotes the
equivalent neutron energy ($U$ can be the excitation energy as well). Then,
the identity of the branching ratios shown in Figs. 3 and 4 can be expressed
as
\begin{equation}
\label{eq:1}
\frac{B_{i}^{S_{1}}(U,J^{\pi})}{B_{i}^{S_{2}}(U,J^{\pi})}
=\frac{R_{i}^{n_{1}}(U)}{R_{i}^{n_{2}}(U)}
\end{equation}
to the accuracy mentioned above, where
\begin{equation}
\label{eq:2}
R_{i}^{n_{j}} \equiv \frac{\sigma_{i}^{n_{j}}}{\sigma_{R}^{n_{j}}}.
\end{equation}
Relation of the $B^x_i$ and $R^x_i$ $(x=S_j~{\rm or}~n_j)$
are expressed as follows:
\begin{equation}
\label{eq:2a}
R_{i}^{x_{j}}(U)  \equiv \frac{\sum_{J^{\pi}}\sigma^{x_{j}}(U,J^{\pi})\cdot
B_{i}^{x_{j}}(U,J^{\pi})}{\sum_{J^{\pi}}\sigma^{x_{j}}(U,J^{\pi})},
\end{equation}
where $\sigma^{x_{j}}(U,J^{\pi})$ denotes the formation cross section 
of $J^\pi$ states in reaction $x_j$ including the factor of $(2J+1)$.
Equation
(\ref{eq:1}) can be rewritten as
\begin{equation}
\label{eq:3}
B_{i}^{S_{1}}(U,J^{\pi})=B_{i}^{S_{2}}(U,J^{\pi})\cdot\frac{R_{i}^{n_{1}}%
(U)}{R_{i}^{n_{2}}(U)}.
\end{equation}
Then, the  decay probability for reaction $i$ in surrogate $S_{1}$ measurement,
$R_{i}^{S_{1}}$, can be written as
\begin{eqnarray}
\label{eq:4}
R_{i}^{S_{1}}(U)  & = &\frac{\sum_{J^{\pi}}\sigma^{S_{1}}(U,J^{\pi})\cdot
B_{i}^{S_{1}}(U,J^{\pi})}{\sum_{J^{\pi}}\sigma^{S_{1}}(U,J^{\pi})} \nonumber \\
&=&\frac
{\sum_{J^{\pi}}\sigma^{S_{1}}(U,J^{\pi})\cdot B_{i}^{S_{2}}(U,J^{\pi}
)\cdot\frac{R_{i}^{n_{1}}(U)}{R_{i}^{n_{2}}(U)}}{\sum_{J^{\pi}}\sigma^{S_{1}
}(U,J^{\pi})}\nonumber\\
& = &\frac{R_{i}^{n_{1}}(U)}{R_{i}^{n_{2}}(U)}\cdot\frac{\sum_{J^{\pi}}
\sigma^{S_{1}}(U,J^{\pi})\cdot B_{i}^{S_{2}}(U,J^{\pi})}{\sum_{J^{\pi}}
\sigma^{S_{1}}(U,J^{\pi})}.
\end{eqnarray}
Since the 2 surrogate reactions $S_{1}$ and $S_{2}$ are assumed to be carried
out for a pair of nuclei having similar mass and structure, the distribution 
of the formation
cross section $\sigma^{S_{1}}(U,J^{\pi})$ will be fairly close to that of 
$\sigma^{S_{2}}(U,J^{\pi})$ if the nuclear structure and reaction
mechanisms are similar to each other. We can write this similarity as
$\sigma^{S_{1}}(U,J^{\pi}) = \alpha \sigma^{S_{2}}(U,J^{\pi})$,
where the symbol $\alpha$ denotes a constant 
such as the kinematical factor.  
If the dependence of $\alpha$ on $J^\pi$ is 
ignorable, Eq. (\ref{eq:4}) reads
\begin{eqnarray}
\label{eq:5}
R_{i}^{S_{1}}(U)&=&\frac{R_{i}^{n_{1}}(U)}{R_{i}^{n_{2}}(U)}\cdot\frac{\sum
_{J^{\pi}} \alpha \sigma^{S_{2}}(U,J^{\pi})\cdot B_{i}^{S_{2}}(U,J^{\pi})}
{\sum_{J^{\pi}} \alpha \sigma^{S_{2}}(U,J^{\pi})} \nonumber \\
&=&\frac{R_{i}^{n_{1}}(U)}{R_{i}
^{n_{2}}(U)}\cdot R_{i}^{S_{2}}(U)
\end{eqnarray}
by definition. 
This equation is equivalent to Eq. (\ref{eq:b}). Since we
know $R_{i}^{n_{2}}(U)$, and we measure $R_{i}^{S_{1}}/R_{i}^{S_{2}}$ in
surrogate ratio method, we can obtain $R_{i}^{n_{1}}$ to
the accuracy mentioned above. This gives an explanation 
of the reason why the surrogate ratio method works.

The essential point in the verification is the equality given in Eq.
(\ref{eq:1}) and equality of the $J^\pi$ spectra of the 2 surrogate
reactions. The latter implies that the $J^\pi$ distributions in 
the surrogate reactions can be
different from those of the neutron-induced reactions.  
What is important is that 2 surrogate reactions should yield equivalent
$J^{\pi}$ distributions.  It can be easily achieved in experiments by 
selecting targets having similar structure and using the same reaction
for the both surrogate reactions.  
However, the difference of the representative spin between the neutron-
induced and surrogate reactions
should not be much larger than about 10 $\hbar$.
We define 
the equality given in Eq. (\ref{eq:1}) as ``{\bf weak Weisskopf-Ewing
condition}". This condition is different from the standard Weisskopf-Ewing
condition, which is written as
\begin{equation}
\label{eq:}
B_{i}^{S_{j}}(U,J^{\pi})=B_{i}^{S_{j}}(U)=R_{i}^{n_{j}}(U).
\end{equation}
If this standard condition is satisfied, we can determine the 
branching ratios by the
surrogate absolute method. Unfortunately, it is not the
case for the reactions investigated in this paper, especially it is a poor
assumption for the capture reaction as shown in Fig. 2.

The surrogate ratio method has another advantage over the
absolute method. Since the surrogate method
uses multi-nucleon transfer reactions very often, 
there is a possibility, when the
corresponding neutron energy increases, that the nucleons expected to be
transferred to bound states of the target is actually transferred
to an unbound state, eventually leading to the breakup (or preequilibrium) 
reactions such
as $^{238}$U($t,np$)$^{239}$U instead of expected reaction $^{238}$%
U($t,p$)$^{240}$U. This effect can be also canceled out by the surrogate
ratio method as follows.

Let us denote the bound states as ``Q", and unbound ones as
``P". Since we measure the ejectile (e.g., $p$), the
production cross section of it 
contains transitions to both the Q- and P-states of the residual
nuclei. On
the contrary, the true decay occurs only via the Q-states. Therefore, the
decay probabilities measured in the surrogate method in the presence of
breakup reaction, $R_{i}^{S_{1}}(P+Q)$, can be written as
\begin{equation}
\label{eq:8}
R_{i}^{S_{1}}(P+Q)=
\frac{\sum_{J^{\pi}}\hat{Q}\sigma^{S_{1}}(U,J^{\pi})\cdot
B_{i}^{S_{1}}(U,J^{\pi})}{\sum_{J^{\pi}}(\hat{P}+\hat{Q})\sigma^{S_{1}}(U,J^{\pi})}
\leq R_{i}^{S_{1}}(U)
=
\frac{\sum_{J^{\pi}}\hat{Q}\sigma^{S_{1}}(U,J^{\pi})\cdot
B_{i}^{S_{1}}(U,J^{\pi})}{\sum_{J^{\pi}}\hat{Q}\sigma^{S_{1}}(U,J^{\pi})},
\end{equation}
where the $\hat{Q}$ and $\hat{P}$ denote fractions of transitions 
to the Q- and
P-states, respectively, and $\hat{P}+\hat{Q}$=1. The same is true for the
$S_{2}$ reaction Therefore, the ratio of the measured surrogate reaction 
ratios reads
\begin{eqnarray}
\label{eq:9}
\frac{R_{i}^{S_{1}}(P+Q)}{R_{i}^{S_{2}}(P+Q)}  &=& 
\frac{
\frac{\sum_{J^{\pi}}\hat{Q}\sigma^{S_{1}}(U,J^{\pi})\cdot B_{i}^{S_{1}}(U,J^{\pi})}{\sum_{J^{\pi}}(\hat{P}+\hat{Q})\sigma^{S_{1}}(U,J^{\pi})}}
{\frac{\sum_{J^{\pi}}\hat{Q}\sigma^{S_{2}}(U,J^{\pi})\cdot B_{i}^{S_{2}}(U,J^{\pi})}{\sum_{J^{\pi}}(\hat{P}+\hat{Q})\sigma^{S_{2}}(U,J^{\pi})}} \nonumber \\
&=&\frac{R_{i}^{n_{1}}(U)}{R_{i}^{n_{2}}(U)}\cdot\frac{\sum_{J^{\pi}}\hat
{Q}\sigma^{S_{2}}(U,J^{\pi})\cdot B_{i}^{S_{2}}(U,J^{\pi})}{\sum_{J^{\pi}}
\hat{Q}\sigma^{S_{2}}(U,J^{\pi})\cdot B_{i}^{S_{2}}(U,J^{\pi})} \nonumber \\
&=&\frac{R_{i}^{n_{1}}(U)}{R_{i}^{n_{2}}(U)},
\end{eqnarray}
where the weak Weisskopf-Ewing condition (Eq. (\ref{eq:3})) 
and proportionality 
of $\sigma^{S_{1}}(U,J^{\pi})$ and $\sigma^{S_{2}}(U,J^{\pi})$ were employed. 
Therefore, the
surrogate ratio method has a capability to work even when
breakup (or preequilibrium) reaction occurs. 

Even though the derivation here is qualitative, it was enough to assume that
the ratios of $\hat{P}$ and $\hat{Q}$ to be the same for the 2 
surrogate reactions used in the ratio method.  This can be satisfied 
if the breakup mechanisms are the same, which is a reasonable assumption.
Again, it must be noted that we do not need to understand the breakup 
reaction mechanism itself, which is a formidable task, but just require 
them to be the same for the 2 reactions employed in the 
ratio method.  It can be easily
verified experimentally by observing the spectra of emitted particles.
This may explain the reason why the  
ratio method worked to measure the $^{236}$U($n,f$) cross
section for energies above several MeV as reported by Lyles \textit{et
al.}\cite{lyles07} where the 2nd and 3rd chance fission occur, which 
corresponds to the condition that the breakup reaction can occur in the 
surrogate method. 

\section{Concluding remarks}

We have investigated the condition that the surrogate reaction should work. It
was found that the surrogate absolute method will give a
marginal result for fission cross sections but it seems to be 
hopeless to apply it for
the capture cross section measurements. On the contrary, 
it was shown that, without assuming any specific
(schematic) spin parity distributions, the surrogate
 ratio method has a high potential to determine 
neutron fission and capture cross sections. The achievable accuracy would be
around 3$\sim$5 \% for the fission and 10 \% for the capture cross sections 
under the
condition investigated in this work (up to difference of spin values of between
neutron-induced and surrogate reactions of
around 10 $\hbar$) for nuclei in Uranium region at around 2.5 to 5 MeV. 
The success is brought by the
weak Weisskopf-Ewing condition, namely, $J^{\pi}$ by $J^{\pi}$ 
convergence of the branching ratios and their coincidence to the
neutron reaction ratio, defined in this work.   Furthermore, 
it is important to select a pair of nuclei, one of which is the reference
nucleus, having similar properties
so that the excitation spectra of various $J^\pi$ states can be 
considered almost equivalent.  These conditions are the basis
for the surrogate ratio method to work. Furthermore, it 
was shown to be rather robust even breakup reaction
occurs.  This was shown again without assuming any breakup reaction 
mechanisms.   Altogether, the surrogate ratio method was proved
to be a useful method to determine neutron cross sections for
which the direct measurements using neutrons are not possible.  
Generally speaking, however, application of the surrogate method must
be done with a caution.  It will be very sensitive to the spin and 
parity of the decaying nucleus at low energies since transitions
to discrete levels, which differ nucleus to nucleus,
 occupy a dominant part of the decay branch there.  This is the reason
why the weak Weisskopf-Ewing condition tends to be violated at 
lower energies.

It must be also noted that we use a standard 
Hauser-Feshbach calculation using
models and parameters adjusted to reproduce neutron cross sections, 
but the results may have some dependence on
them. Such a dependence, however, is expected also to be small in the
surrogate ratio method, since many factors in models and
parameters can cancel out in the ratio quantities.   

\begin{acknowledgments}
The authors are grateful to Drs. A. Iwamoto, 
K. Nishio, S. Hashimoto and Y. Aritomo 
for fruitful discussions. 
Present study is the result of 
``Development of a Novel Technique for Measurement
of Nuclear Data Influencing the Design of Advanced Fast Reactors"
entrusted to Japan Atomic 
Energy Agency (JAEA) by the
Ministry of Education, Culture, Sports, Science and Technology of Japan
(MEXT).
\end{acknowledgments}


\newpage

\begin{figure}
\includegraphics{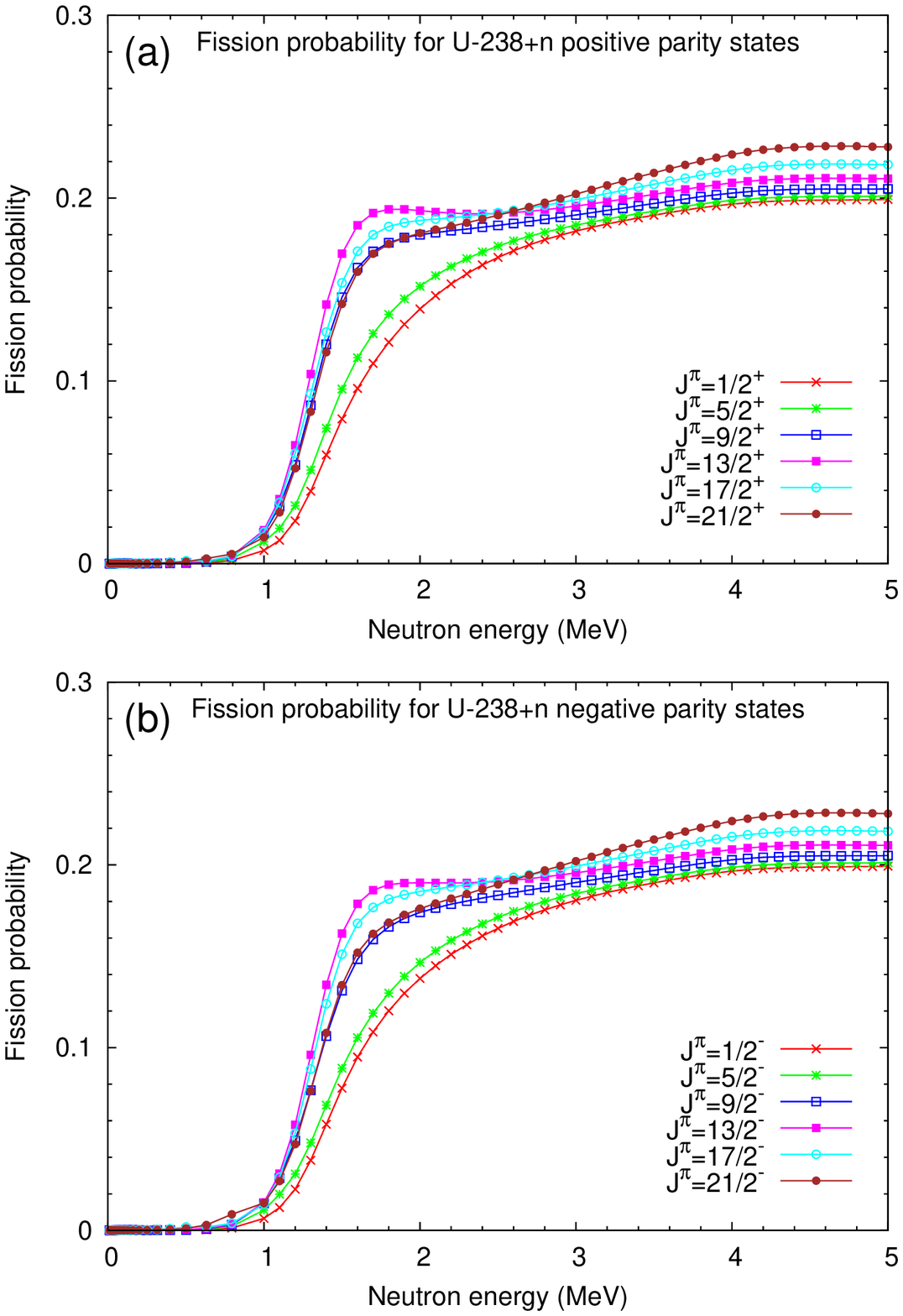}
\caption{\label{figure1}(Color online) Decay probabilities (branching ratios)
to the fission channel from various $J^{\pi}$ states of $^{239}$U. (a)
: positive parity states, (b): negative parity states}
\end{figure}

\begin{figure}
\includegraphics{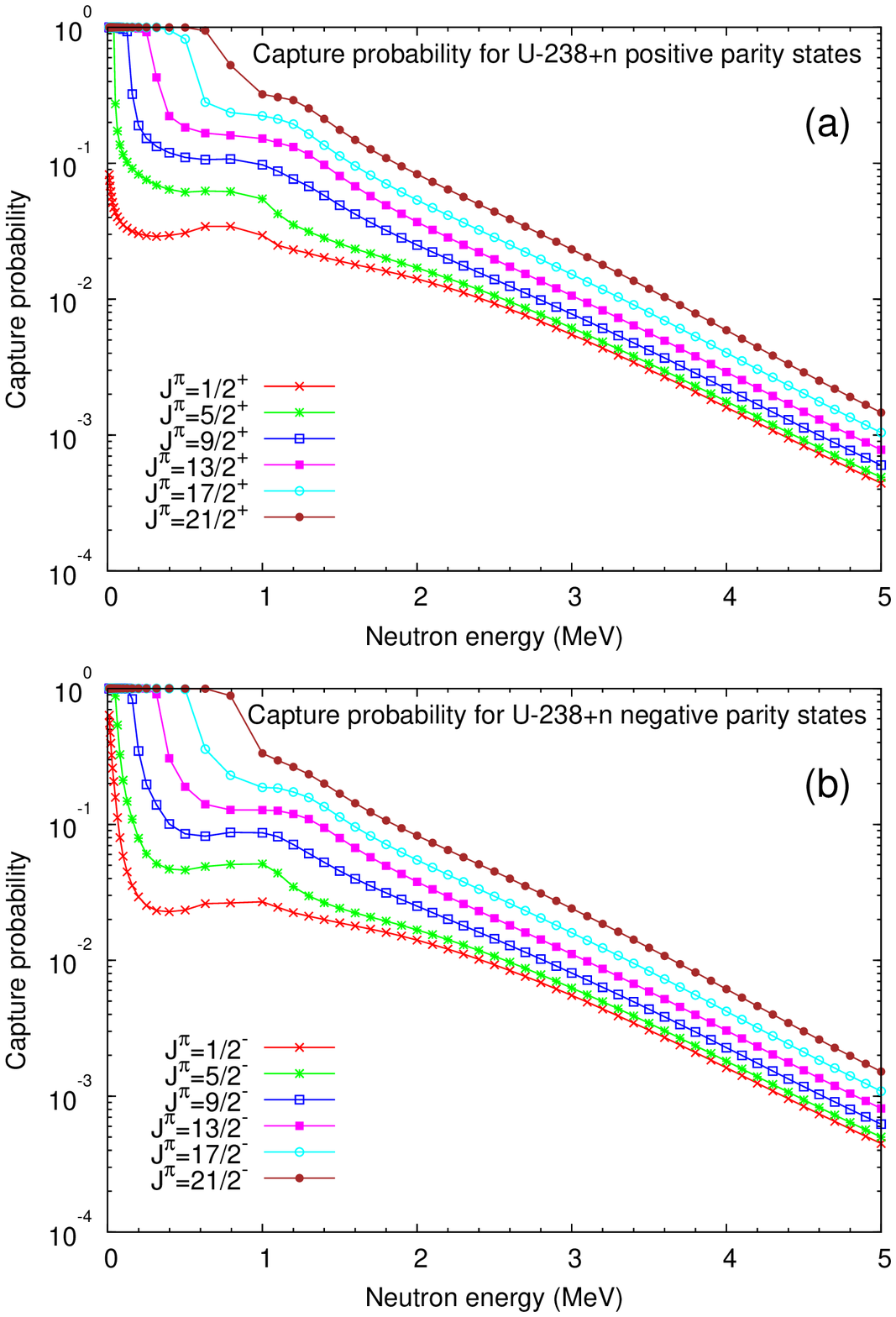}
\caption{\label{figure2}(Color online) Decay probabilities (branching ratios)
to the capture channel from various $J^{\pi}$ states of $^{239}$U. 
(a): positive parity states, (b): negative parity states}
\end{figure}

\begin{figure}
\includegraphics{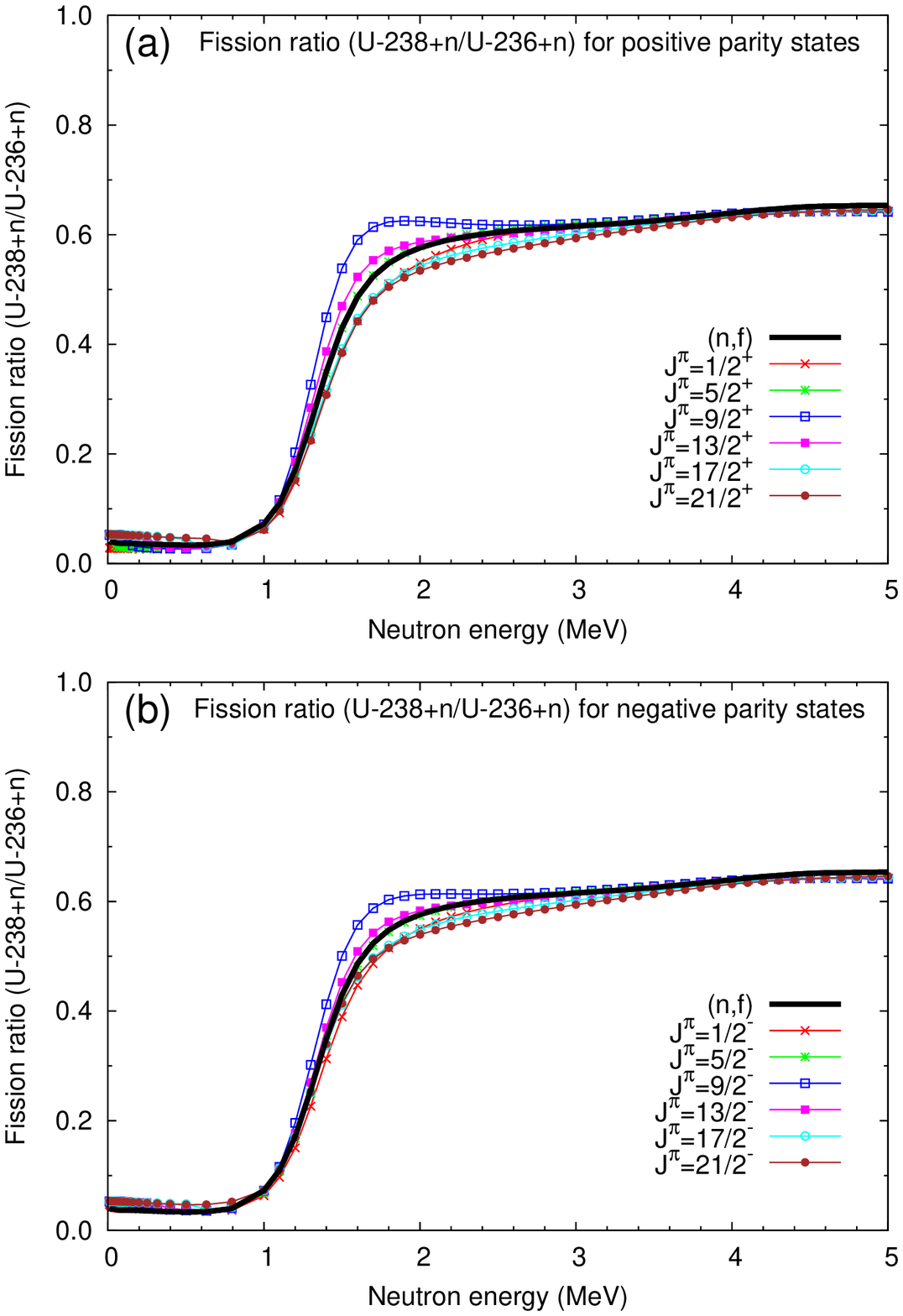}
\caption{\label{figure3}(Color online) Ratios of decay probabilities
(branching ratios) to the fission channel from various $J^{\pi}$ states of
$^{239}$U and $^{237}$U. (a): positive parity states, (b):
negative parity states}
\end{figure}

\begin{figure}
\includegraphics{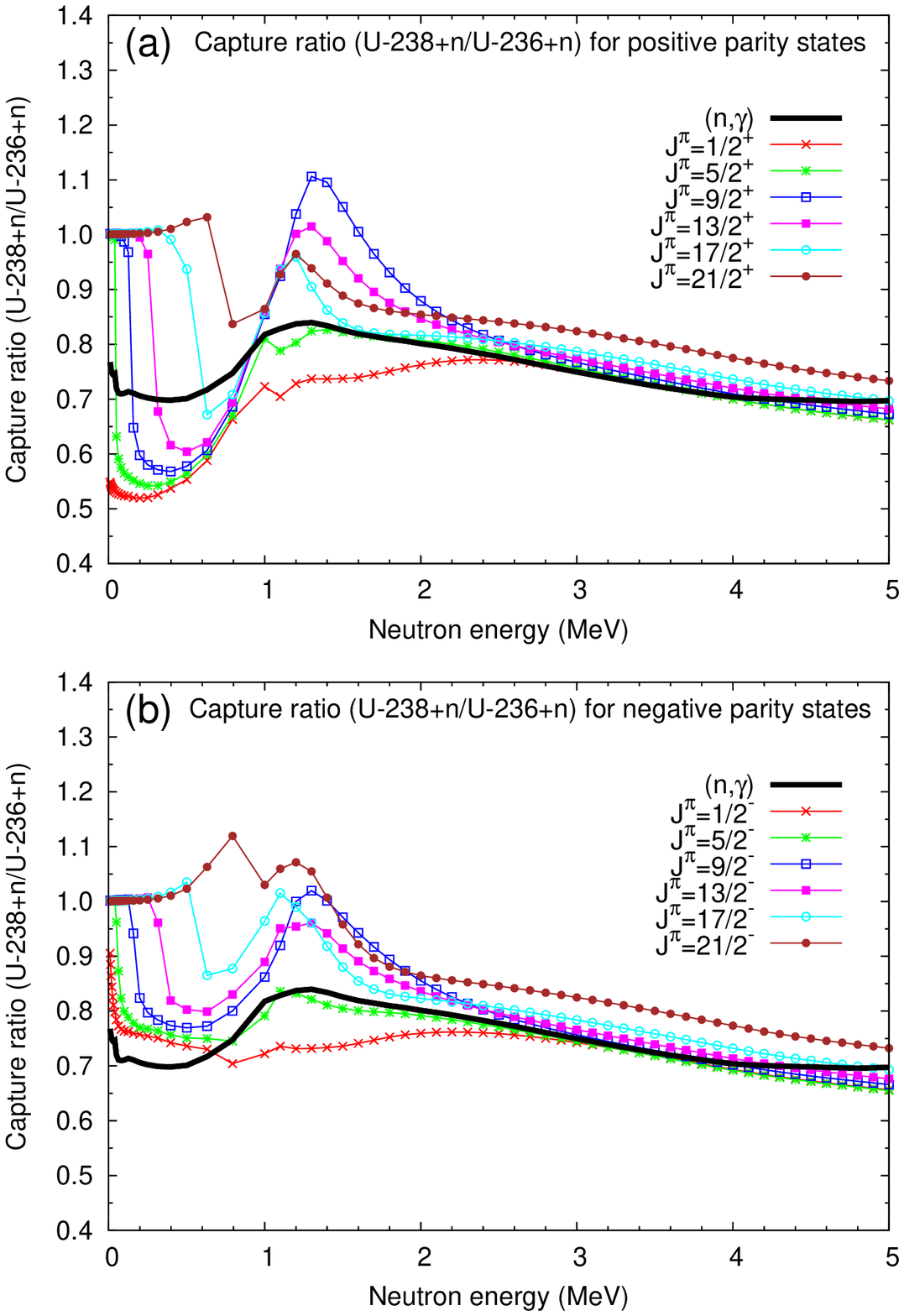}
\caption{\label{figure4}(Color online) Ratios of decay probabilities
(branching ratios) to the capture channel from various $J^{\pi}$ states of
$^{239}$U and $^{237}$U. (a): positive parity states, (b):
negative parity states}
\end{figure}

\begin{figure}
\includegraphics{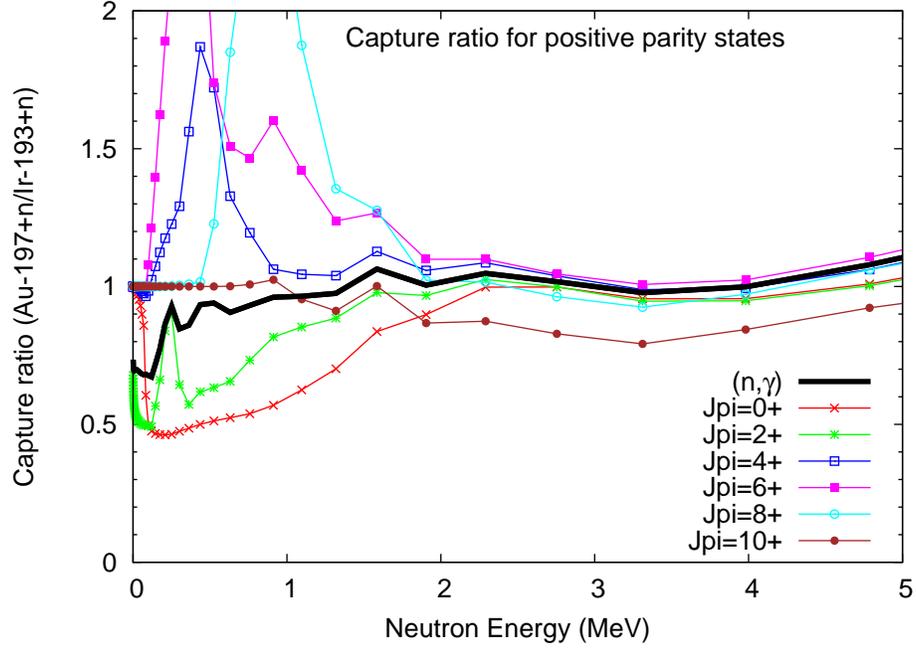}
\caption{\label{figure5}(Color online) Ratios of decay probabilities
(branching ratios) to the capture channel from various $J^{+}$ states of
$^{198}$Au and $^{194}$Ir as a function of corresponding neutron energy
in the case they are produced by neutron-induced reactions. 
Similar results were obtained for negative
parity states.}
\end{figure}

\end{document}